# Active control of near-field radiative heat transfer by coating-twisting method

**Mingjian He,**[1,2] **Hong Qi,**[1,2,*] **Yatao Ren,**[1,2] **Yijun Zhao,**[1] **Mauro Antezza**[3,4,**]

[1] *School of Energy Science and Engineering, Harbin Institute of Technology, Harbin 150001, P. R. China*
[2] *Key Laboratory of Aerospace Thermophysics, Ministry of Industry and Information Technology, Harbin 150001, P. R. China*
[3] *Laboratoire Charles Coulomb (L2C), UMR 5221 CNRS-Université de Montpellier, F-34095 Montpellier, France*
[4] *Institut Universitaire de France, 1 rue Descartes, F-75231 Paris, France*
*Corresponding author: qihong@hit.edu.cn
**Corresponding author: mauro.antezza@umontpellier.fr

**In this letter, active control of near-field radiative heat transfer (NFRHT) between two isotropic materials is realized by a coating-twisting method. The two slabs are coated with graphene gratings, and then the NFRHT can be not only enhanced, but also weakened, by tuning the twisted angle between the two gratings. The physical mechanism is attributed to the modes coupled by the graphene gratings and the isotropic material, which can vary with the twisted angle. The proposed method is also applicable for other kinds of anisotropic films, and may provide a way to realize high-precision nanoscale thermal management, nimble thermal communications and thermal switch.** © 2019 Optical Society of America

http://dx.doi.org/10.1364/OL.99.099999

The radiative heat transfer between two objects can exceed the far-field limit of the Planck's blackbody theory by orders of magnitude, when they are separated by a vacuum gap smaller than the characteristic wavelength of the thermal radiation [1]. In recent years, numerous unique phenomena [2-7] and nanoscale thermal devices [8-13] have been found and proposed in the near-field radiative heat transfer (NFRHT). Among them, to strengthen the NFRHT has always been a topic of general interest due to the pressing needs in heat transfer enhancement in dense electronic devices [14, 15]. However, in some specific cases and nano-micro devices, like micro-electro-mechanical system, a flexible method which can both enhance and weaken the NFRHT at the same time is essential. For natural anisotropic materials, such as hexagonal boron nitride bulk [16, 17], and black phosphorus film [18], the NFRHT between them is found to be tunable by changing the twisted angle between the upper and bottom objects. However, the twisting-tunable NFRHT between two-dimensional materials mainly focuses on the suspended films, without considering the effect of substrate [7, 18], which lacks significance in experimental physics. As for the isotropic materials, due to the uniform in-plane characteristics in different directions [19], the only method to tune the NFRHT between them is to convert them to nanostructured metamaterials, like gratings [20], or nanoholes[21], which needs high manufacturing precision and stability. Until now, there is no method to control the NFRHT between isotropic materials at the fixed vacuum separation without destructing the structure.

In this letter, we cover the isotropic materials with graphene gratings, which are easy for processing in practical preparation process of graphene. Accordingly, the modes of graphene gratings strongly interact with the modes of substrate to form coupled modes, which can vary with the twisted angle. Hence, the NFRHT in the system can be not only enhanced, but also weakened compared to that between bare slabs, which provides the possibility for radiative thermal switch and precise nanoscale thermal management.

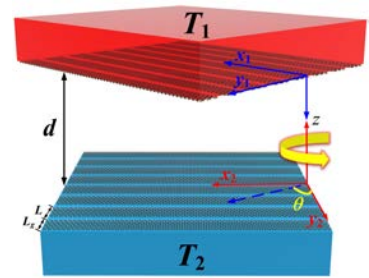

Fig. 1. NFRHT between two isotropic slabs coated by graphene gratings, with periodicity and stripe width $L$ and $L_g$, respectively. The twisted angle between the main axes of the two gratings is $\theta$. The two slabs are separated by vacuum gap $d$ and kept at temperature $T_1$ and $T_2$.

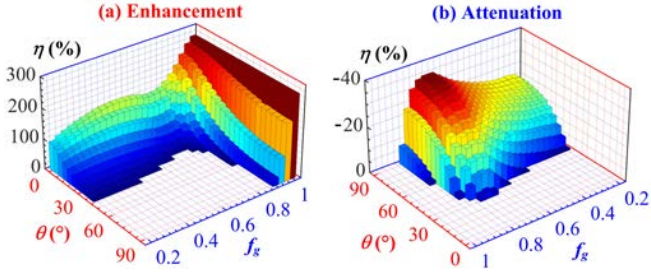

Fig. 2. For different twisted angle $\theta$ and graphene filling factor $f_g$, the (a) enhancement and (b) attenuation effect are demonstrated by factor $\eta$, which is normalized to the NFRHT between two bare slabs without graphene gratings. The parameters are selected as $L$=20 nm, $d$=60 nm, and $T$=300 K respectively.

The proposed coating-twisting method is illustrated by Fig. 1, where two isotropic slabs (made of SiC [22] in the calculations) are separated by a vacuum gap with distance $d$ and coated with graphene gratings. The periodicity and stripe width of the gratings are $L$ and $L_g$, respectively. The twisted angle is defined by the main axes of the two gratings, and can be tuned by mechanical rotation. To characterize the heat transfer in the system, the radiative heat transfer coefficient is defined to evaluate the NFRHT as

$$h = \lim_{\Delta T \to 0} \frac{\Delta \Phi}{\Delta T} = \frac{1}{8\pi^3} \int_0^\infty \hbar\omega \frac{\partial n}{\partial T} d\omega \int_0^{2\pi} \int_0^\infty \xi(\omega,\kappa,\phi) \kappa d\kappa d\phi \quad (1)$$

where $\Delta\Phi$ is the net radiative heat flux and $\Delta T = T_1 - T_2$ is the temperature difference between the two slabs. $\hbar$ is Planck's constant divided by $2\pi$ and $n=[\exp(\hbar\omega/k_BT)-1]^{-1}$ denotes the mean photon occupation number. $\xi(\omega,\kappa,\phi)$ is the energy transmission coefficient, which reads [23]

$$\xi(\omega,\kappa,\phi) = \begin{cases} \text{Tr}\left[\left(\mathbf{I}-\mathbf{R}_2^\dagger\mathbf{R}_2\right)\mathbf{D}\left(\mathbf{I}-\mathbf{R}_1\mathbf{R}_1^\dagger\right)\mathbf{D}^\dagger\right], \kappa < \kappa_0 \\ \text{Tr}\left[\left(\mathbf{R}_2^\dagger-\mathbf{R}_2\right)\mathbf{D}\left(\mathbf{R}_1-\mathbf{R}_1^\dagger\right)\mathbf{D}^\dagger\right]e^{-2|\kappa_z|d}, \kappa > \kappa_0 \end{cases} \quad (2)$$

where $\kappa$ and $\phi$ are the surface-parallel wave vector and the azimuthal angle. $\kappa_0=\omega/c$ and $\kappa_z = \sqrt{\kappa_0^2 - \kappa^2}$ are the wave vector and tangential wave vector in vacuum. $\mathbf{D}=\left(\mathbf{I}-\mathbf{R}_1\mathbf{R}_2 e^{2i\kappa_z d}\right)^{-1}$ is the Fabry-Perot-like denominator matrix and $\mathbf{R}_i$ ($j$=1, 2) is the 2×2 reflection coefficient matrix for the $i$-th interface, with the elements [24, 25]

$$\mathbf{R} = \begin{bmatrix} \frac{(\rho_a+1)(1-\rho_c)+\rho_b\rho_d}{(\rho_a+1)(\rho_c+1)-\rho_b\rho_d} & \frac{2\rho_b}{(\rho_a+1)(\rho_c+1)-\rho_b\rho_d} \\ \frac{-2\rho_d}{(\rho_a+1)(\rho_c+1)-\rho_b\rho_d} & \frac{(\rho_a-1)(1+\rho_c)-\rho_b\rho_d}{(\rho_a+1)(\rho_c+1)-\rho_b\rho_d} \end{bmatrix} \quad (3)$$

where $\rho_a = (\sigma_{xx}/\omega\varepsilon_0 + \varepsilon_i/\kappa_{z,i})\kappa_z$, $\rho_b = \sqrt{\mu_0/\varepsilon_0}\sigma_{xy}$, $\rho_c = (\sigma_{yy}\omega\mu_0 + \kappa_{z,i})/\kappa_z$ and $\rho_d = \sqrt{\mu_0/\varepsilon_0}\sigma_{yx}$. $\varepsilon_0$ and $\mu_0$ are the permittivity and permeability of vacuum, respectively. $\kappa_{z,i} = \sqrt{\varepsilon_i\kappa_0^2 - \kappa^2}$ is the tangential wave vector in the substrate with the relative permittivity $\varepsilon_i$. The conductivities $\sigma_{xx}$, $\sigma_{yy}$, $\sigma_{xy}$, and $\sigma_{yx}$ are modified by the effective conductivity for two twisted graphene gratings [26]

$$\begin{aligned}\sigma_{xx} &= \sigma_\parallel \cos^2\phi_i + \sigma_\perp \sin^2\phi_i \\ \sigma_{xy} &= \sigma_{yx} = (\sigma_\parallel - \sigma_\perp)\sin 2\phi_i/2 \\ \sigma_{yy} &= \sigma_\parallel \sin^2\phi_i + \sigma_\perp \cos^2\phi_i\end{aligned} \quad (4)$$

where the effective conductivities $\sigma_\parallel = f_g\sigma_g$ and $\sigma_\perp = \sigma_g\sigma_c/(f_g\sigma_c + f_c\sigma_g)$ are the conductivities along and across the main axes of the graphene gratings. $f_g=L_g/L$ is the graphene filling factor, and $f_c=1-f_g$ is the filling factor of the free space between adjacent stripes. As for the conductivity of graphene $\sigma_g$, it can be written as a sum of an intraband and an interband contribution, given by [27]. $\sigma_c = -\frac{i\omega\varepsilon_0 L(\varepsilon_0+\varepsilon_i)}{\pi}\ln\left[\csc\left(\frac{\pi f_c}{2}\right)\right]$ [28] is an equivalent conductivity associated with the coupling between adjacent graphene stripes. It should be mentioned that, when the gratings are twisted, the relationship of the azimuthal angles for the two gratings is $\phi_2=-\phi_1+\theta$[21].

To quantitatively study the modulation ability of proposed method, the modulation factor

$$\eta = \left[h(\theta,f_g) - h_{f_g=0}\right]/h_{f_g=0} \quad (5)$$

where $h(\theta,f_g)$ is the radiative heat transfer coefficients of the system shown in Fig. 1 with specific twisted angle $\theta$ and graphene filling factor $f_g$, and $h_{f_g=0}$ is the coefficient without graphene gratings. In Fig. 2, the enhancement and attenuation effect driven by twisting are demonstrated with $\eta$ at different $f_g$. The chemical potential of graphene is set as $\mu$=0.5 eV and the other parameters are selected as $L$=20 nm, $d$=60 nm, and $T$=300 K respectively. The results show clearly that the NFRHT can be actively and largely enhanced or weakened by mechanical rotation at around $f_g$=0.2-0.9. Specifically, at $f_g$=0.8, the NFRHT can be enhanced by about 150% and weakened by about 30% compared to the NFRHT of bare slabs.

To study the underlying physical mechanism of the strong modulation ability, in Fig. 3(a), we have given the spectral radiative heat transfer coefficient at different twisted angles and that without graphene gratings. The spectral results show that there are three dominant modes in the system and they obviously vary with the twisted angle. In Figs. 3(b)-(d), the energy transmission coefficients for the specific twisted angles are given to reveal the modes in the system. The contours are quite different from those of bare SiC system or SiC slabs covered by graphene sheets [29]. For aligned gratings in Fig. 3(b), the three dominant modes are labeled with the dashed lines in different colours at $\omega$=0.4×$10^{14}$rad/s, 1.84×$10^{14}$rad/s, and 2.4×$10^{14}$rad/s. To explore the nature of the three different modes, the anisotropic energy transmission coefficients for aligned and orthogonal gratings are given in Fig. 4 at the specific frequencies. The dispersion relations for the upper and bottom surface characteristics, which are calculated by zeroing the denominator of Eq. (3), are denoted by the red and green-dashed lines, respectively. The results reveal that the graphene hyperbolic surface plasmon polaritons (HSPPs) modes appears at low frequencies, which exist widely in graphene grating systems [30]. Moreover, at $\omega$=1.84×$10^{14}$rad/s near the frequency of SiC surface phonon polaritons (SPhPs), another hyperbolic modes appear. Due to the fact that the slabs are made of isotropic materials, the anisotropic-hyperbolic modes are attributed to the graphene gratings. It means that the graphene can couple with the SiC to form coupled SPhPs-HSPPs modes, which differ from the two isolated modes. The coupled SPhPs-HSPPs modes have wider frequency range than pure SiC SPhPs, which is clearly demonstrated in Fig. 3. The interesting phenomenon is mainly attributed to the characteristics of graphene SPPs, of which the modes tends to occupy a wide frequency region [31].

Another interesting phenomenon is that a branch of modes appears at frequencies higher than $2\times10^{14}$rad/s, which has never been observed in grahene grating systems, SiC systems or SiC slabs covered by graphene sheets. The anisotropic energy transmission coefficient contour implies that the modes are elliptic modes. In magnetic-tunable graphene grating system, we have found elliptic graphene SPPs modes for the first time [25]. However, the elliptic modes are caused by the combined effect of quantum Hall regime of magneto-optical graphene and the grating effect. For graphene gratings without magnetic fields, the SPPs modes are always hyperbolic and occupy a continuous frequency region [30]. Therefore, the elliptic modes observed in this system should be induced by the coupled effect of the substrate material, graphene SPPs and the grating effect. As $\theta$ increases, the $h(\omega)$ and $\zeta(\omega,\kappa)$ in Fig. 3 show that the three modes are all weakened and shrink to narrower frequency region. The dashed lines together with the arrows in Figs. 3(c) and 3(d) prove the weakened and shrinkage behavior. Moreover, it can be seen from the anisotropic energy transmission coefficients $\zeta(\omega, \kappa, \phi)$ in Fig. 4 that for aligned gratings, the dispersion relations for the upper and bottom surface characteristics match perfectly due to the zero-$\theta$. Nevertheless, for orthogonal gratings, the 90°-$\theta$ induces great mismatch between the the upper and bottom surface characteristics. As can be seen from the $\zeta(\omega,\kappa,\phi)$ in Fig. 4 that, the energy transmission coefficients reach maximum near the intersections of the two curves determined by the dispersion relations. Hence, the three modes all decay when the twisted angle increases, which is the key to realize the active control of the NFRHT between the two isotropic slabs.

In Fig. 5, the dispersion relations of the two surface characteristics are given as a function of frequencies for two orthogonal gratings. The red and green curves at each $\omega$ are symmetrical in $\kappa_x$-$\kappa_y$ plane and they are in 90° rotationally symmetric due to the twist. The dispersion relations of graphene HSPPs modes in Fig. 5(a) show that as $\omega$ increases, the two pairs of hyperbolas both shrink to be narrower and then they have no intersections any more when $\omega$ is higher than $5\times10^{13}$rad/s. Due to the fact that the modes are excited when the the dispersion relations of the two interfaces match or cross, the cut-off for graphene HSPPs modes in Fig. 3(d) can be well explained by the evolution of the hyperbolas. The similar phenomenon can also be observed for coupled SPhPs-HSPPs modes when $\omega$ is higher than $1.85\times10^{14}$rad/s, which match well with the cut-off frequency in Fig. 3(d). In Fig. 5(c), the transition from coupled SPhPs-HSPPs modes to coupled elliptic modes is clearly demonstrated near $2.15\times10^{14}$rad/s, where the two pairs of hyperbolas convert to two elliptic curves. By analyzing the dispersion relations in the $\omega$ space, it is confirmed that when the twisted angle increases, the converge area of the dispersion relations shrinks and tend to occupy narrower $\omega$-region, and then the modes are weakened accordingly. To realize the key function of active control, the graphene HSPPs modes and the two coupled modes are essential. Moreover, all the three modes are dependent on the existence of the graphene gratings and the filling factors of them.

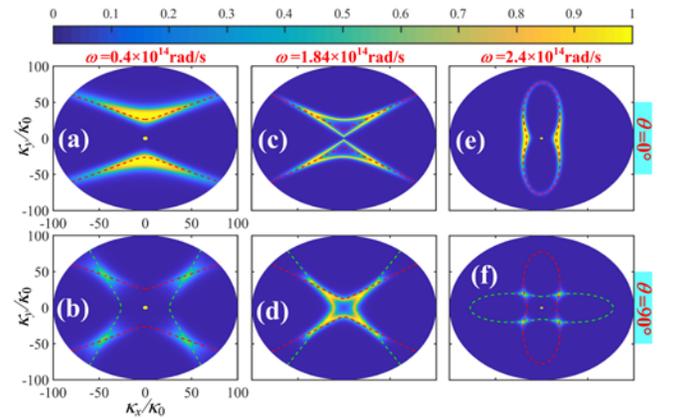

Fig. 4. Anisotropic energy transmission coefficients at $\theta$=0°(the first line) and $\theta$=90°(the second line) for graphene HSPPs modes (the first column), coupled SPhPs-HSPPs modes (the second column) and coupled elliptic modes (the third column). The dispersion relations for the upper and bottom surface characteristics are denoted by the red and green-dashed lines, respectively.

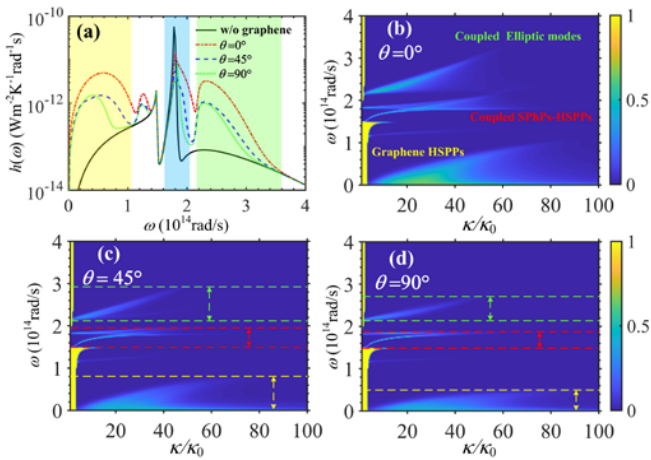

Fig. 3. (a) The spectral radiative heat transfer coefficients for different $\theta$ with filling factor $f_g$=0.8 and that without graphene gratings. Energy transmission coefficients for (b) $\theta$=0°, (c) $\theta$=45° and (d) $\theta$=90°. The graphene HSPPs modes, coupled SPhPs-HSPPs modes and coupled elliptic modes are labeled by dashed lines.

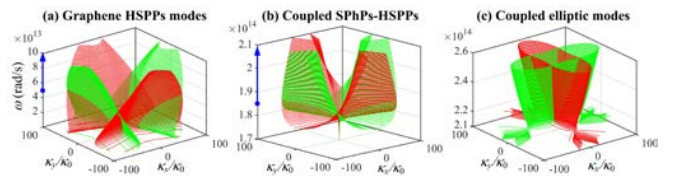

Fig. 5. For orthogonal gratings $\theta$=90°, the dispersion relations of the two surface characteristics at different frequencies.

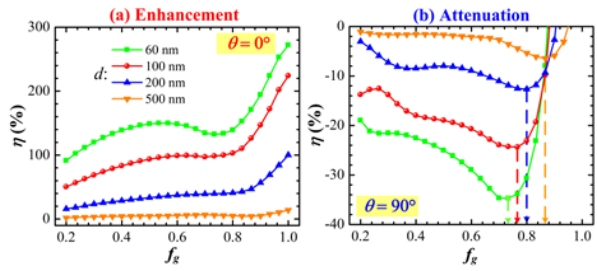

Fig. 6. For different vacuum gap distance, the (a) enhancement ($\theta=0°$) and (b) attenuation ($\theta=90°$) factors $\eta$ compared to NFRHT between two pure slabs without graphene gratings at different $f_g$.

It is known that the graphene SPPs can only be excited near the interface of graphene sheet, and decay as the $d$ increases. In Fig. 6, the enhancement and attenuation factors $\eta$ are respectively calculated for $\theta=0°$ and $\theta=90°$ when $d$ equals 60, 100, 200 and 500 nm. The control function gradually breaks down as the vacuum separation enlarges. The similar attenuated phenomenon has also been observed in two twisted SiC gratings system, which accounts for the attenuation of SPhPs of SiC [20]. Hence, the decaying trend of modulation function is widely observed in NFRHT system due to the nature of the surface modes. Specifically, for different $d$, the NFRHT can always be enhanced by aligned gratings and be weakened by orthogonal gratings. It means that at specific twisted angle, $\eta$ equals zero and NFRHT can be the same as that between two bare SiC slabs. The interesting phenomenon implies that the proposed coating-twisting method could also act like bare system without graphene gratings, which provides the stealth possibility for covering materials.

In conclusion, we have theoretically proposed an active control method, which can not only enhance, but also weaken the NFRHT between two isotropic slabs at the same time just by mechanical twist. Moreover, at some specific twisted angles, the NFRHT can be the same as that without covering, which means the system also has the function to conceal the covering. The heat transfer can be enhanced by about 150% and weakened by about 30% compared to the NFRHT of bare slabs at specific graphene filling factor. By covering the isotropic slabs with graphene gratings, the substrate material couple with the graphene gratings to form three different modes, i.e., the graphene HSPPs modes, the coupled SPhPs-SPPs modes, and the coupled elliptic mode. All the three modes vary with the matching degree of the two interfaces, which is determined by the twisted angle. As the result that the method proposed in this letter is attributed to the anisotropic characteristics of the covering gratings, the proposed method is also applicable for other kinds of anisotropic films, such as $MoS_2$, black phosphorus and so on. The existence of the substrate also provides significance in experimental physics for the novel kinds of two-dimensional thin films. Featured by its strong practicability, easy implement, novel construction, and simple structure, the proposed coating-twisting method may provide a way to realize high-precision nanoscale thermal management and nimble thermal switch.

**Funding.** National Natural Science Foundation of China (NSFC) (51976044, 51806047)

**Acknowledgment.** The authors acknowledge support from Heilongjiang Touyan Innovation Team Program. M. A. acknowledges support from the Institute Universitaire de France, Paris, France (UE).

**Disclosures.** The authors declare no conflicts of interest.